# Electro-optic and magneto-optic photonic devices based on multilayers photonic structures

Giuseppe M. Paternò[1], Liliana Moscardi[2], Ilka Kriegel[3,4], Francesco Scotognella[1,5]*, Guglielmo Lanzani[1,5]

[1] *Center for Nano Science and Technology@PoliMi, Istituto Italiano di Tecnologia, Via Giovanni Pascoli, 70/3, 20133, Milan, Italy*
[2] *Dipartimento di Chimica, Materiali e Ingegneria Chimica "Giulio Natta", Politecnico di Milano, Piazza Leonardo da Vinci 32, 20133 Milano, Italy*
[3] *Department of Nanochemistry, Istituto Italiano di Tecnologia (IIT), via Morego, 30, 16163 Genova, Italy*
[4] *The Molecular Foundry, Lawrence Berkeley National Laboratory, Berkeley, California 94720, United States*
[5] *Dipartimento di Fisica, Politecnico di Milano, Piazza Leonardo da Vinci 32, 20133 Milano, Italy*

**Abstract.** In this work we describe different types of photonic structures that allow tunability of the photonic band gap upon the application of external stimuli, as the electric or magnetic field. We review and compare two porous 1D photonic crystals: in the first one a liquid crystal has been infiltrated in the pores of the nanoparticle network, while in the second one the optical response to the electric field of metallic nanoparticles has been exploited. Then, we present a 1D photonic crystal made with indium tin oxide (ITO) nanoparticles, and we propose this system for electro-optic tuning. Finally, we describe a microcavity with a defect mode that is tuned in the near infrared by the magnetic field, envisaging a contact-less magneto-optic switch. These optical switches can find applications in ICT and electrochromic windows.

…
e-mail address: francesco.scotognella@polimi.it

**Introduction**

The active tuning of the optical response of photonic crystals is attracting increasing attention in view of several applications, as for example electro-optic switching for information and communication technology, or electrochromic windows for the spectral control of sun irradiation of buildings in order to optimize their heating or the realization of new, color displays. Due to the refractive index modulation in a photonic crystal[1–3], the photonic band gap can be actively tuned either by modifying the refractive index of the components or by changing the pitch of the periodic structure. The electric field is a widely used external stimulus for achieving resonance tuning , as documented by an extensive literature[4]. For example, a broad tunability that covers all the visible spectra have been reported with opals and inverse opals based on poly ferrocenylsilane [5,6]. Alternatively, one can achieve tuning of the photonic band gap by applying an external magnetic field. This is a very interesting approach to make contactless switches, because it does not need electrodes in contact with the optical structure [7–11].

In this work, we compare our recent results on the electric/magnetic tunability of the photonic band gap of 1D photonic crystals. In the first part, we report on the fabrication of a 1D photonic crystal made of alternating layers of silica and zirconia nanoparticles, infiltrated with the liquid crystal mixture E7 [composed of 4-cyano-4'-n-pentyl-biphenyl (5CB), 4-cyano-4'-n-heptyl-biphenyl (7CB), 4-cyano-4'-n-octyl-biphenyl (8CB) and 4-cyano-4'-pentyl-p-terphenyl [12]]; such system has shown a blue shift of the band gap of 8 nm upon application of a voltage of 8 V [13,14]. In the second part, we show a 1D photonic crystal made of silver nanoparticles alternated with titania nanoparticles; such system has shown a blue shift 10 nm with an external voltage of 10 V [15]. While the first system exploits the electro-optic properties of liquid crystals, the second one is based on the change of carrier density in the metals that affects the plasmon resonance spectral position.

Furthermore, we show an ITO/titania nanoparticle 1D photonic crystal with promising electro-optic tunability and, finally, we demonstrate the possibility to achieve a magnetic field driven tunability of 1D photonic microcavity by infiltrating a magnetically active material as $Tb_{53}Ga_5O_{12}$ (TGG).

**Methods**

*Sample preparation*: Porous photonic crystals, as synthetic opals and nanoparticle based multilayers, can be infiltrated with different materials like metals [16,17] and conjugated polymers [18]. he experimental procedure to fabricate the silica/zirconia nanoparticle 1D photonic crystal and its infiltration with E7 liquid crystal mixture is reported in Ref. [13]. Instead, the process to fabricate the silver/titania nanoparticle 1D photonic crystal is reported in Ref. [15].

For the fabrication of the ITO/titania nanoparticle 1D photonic crystals, we suspended the $TiO_2$ nanoparticles (Gentech Nanomaterials, average size 5 nm) in deionized water to obtain a concentration of 5 wt. %. ITO nanoparticles (Gentech Nanomaterials, average size 20-30 nm) were diluted with distilled water to a final concentration of 5 wt. %. The dispersions were sonicated for 120 minutes at room temperature and filtered with a 0.45 μm PVDF filter. The dispersions were then spin-cast on clean glass substrates, which were previously sonicated in isopropanol and acetone for 5 minutes and subjected to an oxygen plasma treatment (10 minutes) to remove organic contaminants and improve wettability. The photonic crystal were fabricated using a spin coater Laurell WS-400- 6NPP-Lite. The rotation speeds for the deposition were 2000 rotations per minute (rpm). After each deposition, the sample were annealed for 10 minutes at 350 °C on a hot plate under the fume hood.

*Optical characterization*: The experimental setups to perform the electric field dependent light transmission for the liquid crystal infiltrated photonic crystal and for the silver/titania photonic crystal are reported in Refs. [13,15] The light transmission spectra of the ITO/titania photonic crystals have been recorded with a Perkin Elmer spectrophotometer Lambda 1050 WB.

*Modeling of the optical properties*: To model the light transmission of the photonic structures we have employed the transfer matrix method, described in [19–23]. We study the arrangement in which light impinges normally the glass substrate, the multilayer and then air. For air and glass the parameters employed in the simulations are the refractive index of glass $n_s$ and the refractive index of air $n_0$. We can write an expression that gives the electric and magnetic fields in air, $E_0$ and $H_0$, with the electric and magnetic fields in glass, $E_m$ and $H_m$:

$$\begin{bmatrix} E_0 \\ H_0 \end{bmatrix} = \prod_{j=1}^{x} M_j \begin{bmatrix} E_m \\ H_m \end{bmatrix} = \begin{bmatrix} m_{11} & m_{12} \\ m_{21} & m_{22} \end{bmatrix} \begin{bmatrix} E_m \\ H_m \end{bmatrix} \tag{1}$$

with $x$ number of layers and $M_j$ the characteristic matrix of the *j*th layer [$j=(1,2,…,x)$]

$$M_j = \begin{bmatrix} \cos(\phi_j) & -\frac{i}{p_j}\sin(\phi_j) \\ -ip_j\sin(\phi_j) & \cos(\phi_j) \end{bmatrix} \tag{2}$$

with $\phi_j$ the phase variation of the light wave passing through the layer. For normal incidence of light $\phi_j = (2\pi/\lambda)n_j d_j$, where $n_j$ is the refractive index of the layer and $d_j$ the layer thickness. $p_j = \sqrt{\varepsilon_j/\mu_j}$ in transverse electric (TE) wave [while $q_j=1/p_j$ replace $p_j$ in transverse magnetic (TM) wave, taking into account that, at normal incidence, the transmission spectra for TE and TM waves are the same.

We can thus write the transmission coefficient

$$t = \frac{2p_s}{(m_{11}+m_{12}p_0)p_s+(m_{21}+m_{22}p_0)} \tag{3},$$

and the transmission

$$T = \frac{p_0}{p_s}|t|^2 \tag{4}.$$

**Results and Discussion**

*Liquid crystal infiltrated 1D porous photonic crystals*: As depicted in the scheme in Figure 1, a 1D porous photonic crystal made of $SiO_2$ and $ZrO_2$ nanoparticles has been fabricated on a indium tin oxide substrate and then sandwiched with an additional ITO substrate on top. Later, the E7 liquid crystal mixture has been infiltrated in the pores among the nanoparticle network.

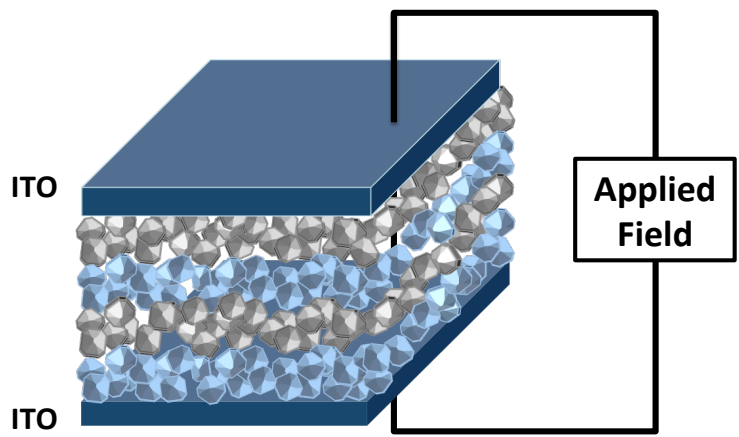

**Figure 1.** Scheme of the 1D porous photonic crystals embedded between two layers of indium tin oxide (ITO) in order to apply an external electric field. We sketched only two bilayers.

When the liquid crystal is embedded in the porous photonic crystals, the refractive index of each layer is due to the suitable combination of the one of silica (or zirconia) and the isotropic refractive index of E7. Modulation comes into play when the voltage is applied, because the refractive index of E7 is then that of the molecules aligned with the electric field (ordinary refractive index) which, in turns, is lower with respect to the isotropic one.

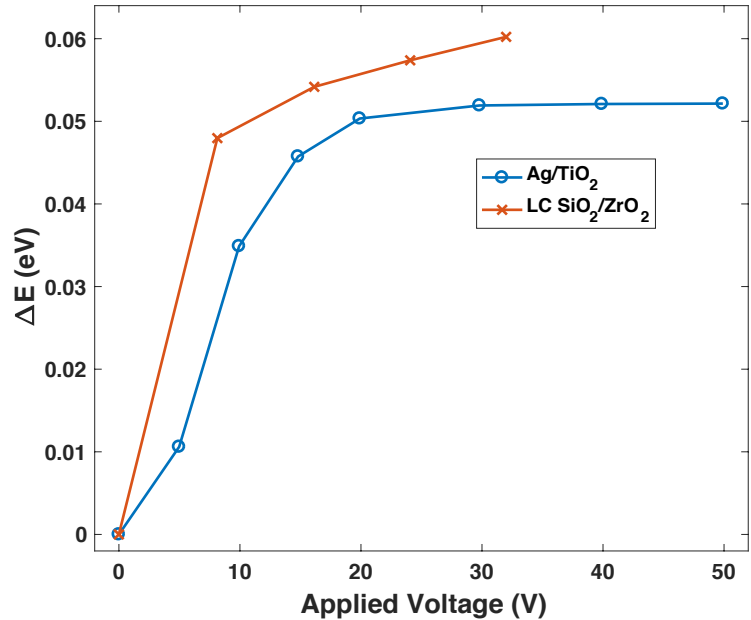

**Figure 2.** Shift of the photonic band gap as a function of the applied voltage. The initial position in wavelength for the liquid crystal infiltrated photonic crystal (LC $SiO_2/ZrO_2$) is 456 nm (2.72 eV), while the one of the $Ag/TiO_2$ is 630 nm(1.97 eV). The voltage for the LC $SiO_2/ZrO_2$ photonic crystal is modulated (square waveform at 1 kHz), while the one for $Ag/TiO_2$ is continuous.

*Tunable silver/titania 1D photonic crystals under electric field*: In a photonic crystal made by silver nanoparticles and titania nanoparticles the active tuning with the electric field (Figure 2, blue curve) is possible because of the possibility to change the number of carrier in the silver nanoparticles, controlling the plasmonic response of this material. In the system the electric field is increasing the number of carriers in the silver nanoparticles. Such change is leading to a blue shift of the plasma frequency $\omega_p = \sqrt{Ne^2/m^*\varepsilon_0}$, with $N$ number of carriers, $e$ the electron charge, $m^*$ the effective mass and $\varepsilon_0$ the vacuum permittivity. By employing the Drude model and the Maxwell-Garnett effective medium approximation to determine the refractive index of the silver nanoparticle layers and we use this refractive index in the transfer matrix theory, we can fit the blue shift of the photonic band gap, finding a good agreement between the experiment and the suggested model [15].

We would like to underline that the first system presented, i.e. the liquid crystal infiltrated photonic crystal, is indeed a liquid/solid state device, while the silver nanoparticle based photonic crystal is a completely solid state device.

*ITO/titania 1D photonic crystals*: We have fabricated 1D porous photonic crystals by employing ITO nanoparticles and titania nanoparticles. In Figure 3a we show the absorption spectra of these photonic crystals for two different rotations per minute (rpm) of the spin coater, thus different layer thicknesses (we used the same rpm values for ITO and titania). In the spectra the photonic band gap at around 500 nm is clearly visible, together with the intense plasmon resonance of ITO in the infrared (we performed the measurement up to 2600 and we observed only the blue tail of the plasmon resonance).

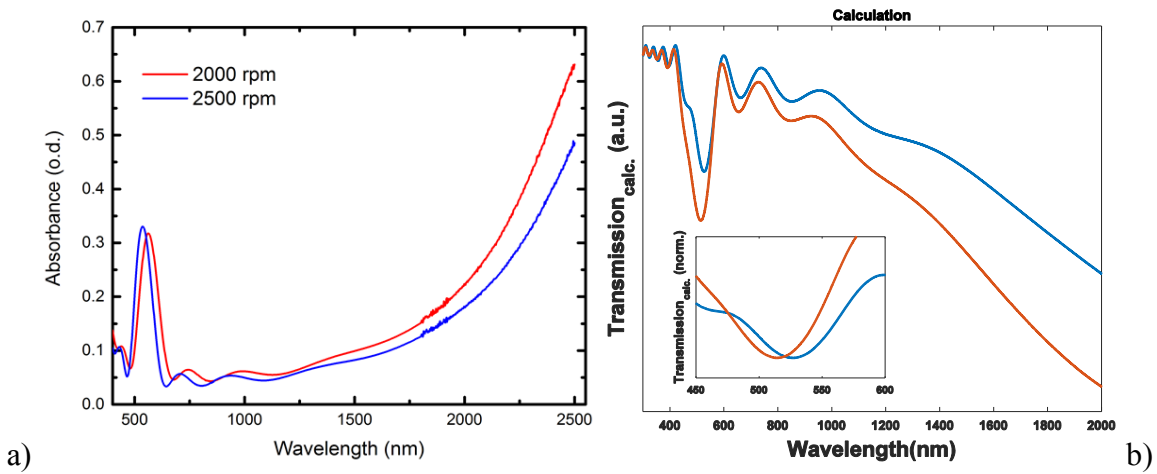

**Figure 3.** (a) Light transmission spectra of ITO/TiO$_2$ photonic crystals as a function of the spin coating speed; (b) Simulation employing the transfer matrix method, the Maxwell-Garnett effective medium theory and the Drude model.

ITO is a very interesting material that belongs to the family of the so-called degenerately doped metal oxides[24]. ITO nanoparticles show a strongly tunable infrared plasmon resonance. For this reason, a photonic crystal that employs ITO nanoparticles is a promising electro-optic switch. In Figure 3b we show that, with a proper change of the plasma frequency and the high frequency of the dielectric function of ITO [25], we observe a clear blue shift of the photonic band gap. In particular, $\omega_{P,1} = 1.24$ eV and $\varepsilon_\infty = 4$ for the blue curve, while $\omega_{P,1} = 1.49$ eV and $\varepsilon_\infty = 3.7$ for the red curve.

*Magneto-optic 1D microcavities*: Recently, very interesting works report the fabrication and characterization of magnetophotonic crystals [26–29]. We suggest here a microcavity with embedded TGG, which is a magnetically active material. We have shown in Ref. [11] the operation of this type of microcavity in the visible range. Here we show the possibility to tune the defect mode of this cavity in the near infrared. TGG was embedded between two photonic crystals made of silica and yttria ($Y_2O_3$). For an accurate analysis of the light transmission through this microcavity, we took into account the dispersion of the refractive index and the Verdet constant of TGG, silica and yttria. The two different refractive indexes for clockwise and counter-clockwise circularly polarized light that propagate through the multilayer structure, with a magnetic field $\vec{B}$ parallel to the direction of propagation of light, can be written in the following way [7]:

$$n_{R,L}(\lambda) = n(\lambda) \pm \frac{V(\lambda)\vec{B}\lambda}{2\pi} \tag{5}$$

with $n(\lambda)$ the wavelength dependent refractive index (taking into account that for all the materials employed we assume $\mu \approx 1$, also for TGG [30]), $V$ the Verdet constant, which is also wavelength dependent.

We fit the experimental data from Ref. [31] to get the dispersion of the refractive index of TGG:

$$n^2_{TGG}(\lambda) - 1 = \frac{2.742\lambda^2}{\lambda^2 - 0.01743} \tag{6}$$

with wavelength in micrometers.

The wavelength and temperature dependent Verdet constant for TGG is take from Ref. [32]:

$$V(\lambda, T) = +\frac{175\lambda_0^2}{\lambda^2 - \lambda_0^2} - \frac{(32 \times 10^4)\lambda_0^2}{(T - T_W)(\lambda^2 - \lambda_0^2)} - \frac{1714}{T - T_W} \tag{7}$$

with wavelength in nm. The Verdet constant is referred to the ceramic material K4 (by Konoshima Chemical, Co. Ltd.), with $\lambda_0$ = 237 nm (wavelength of the absorption resonance) and $T_W$ = 7.6 K (Curie-Weiss temperature).

We have also taken into account the Sellmeier equation and Verdet constant for silica (reported in Ref. [33] and Refs. [34–36], respectively) and for yttria (reported in Ref. [37] and Ref. [38], respectively).

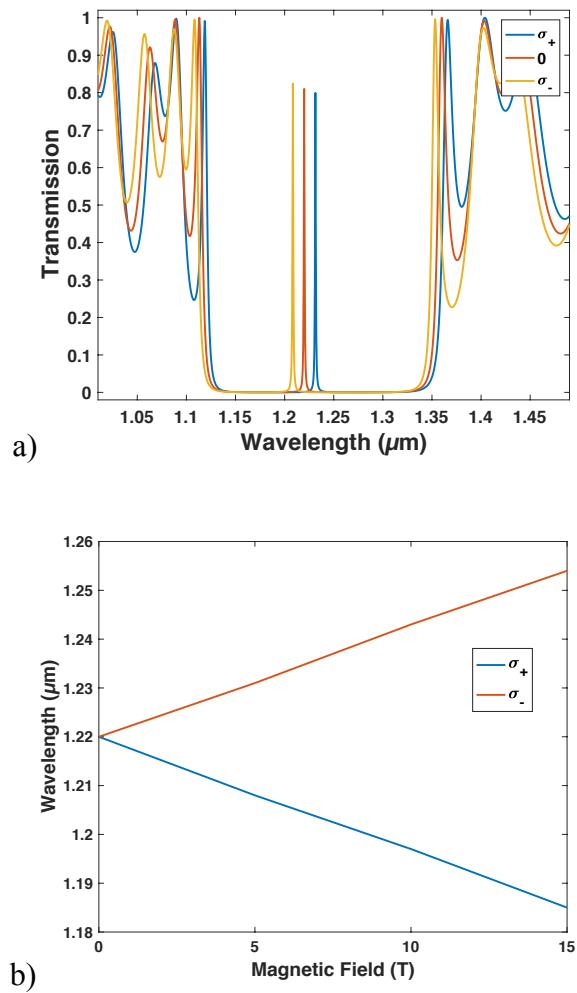

**Figure 4.** a) light transmission spectrum for a TGG microcavity without magnetic field applied (red curve), and with an applied magnetic field of 5 T, at a temperature of 8 K, for the clockwise $\sigma_+$ (blue curve) and the counter-clockwise $\sigma_-$ (dark yellow curve) polarizations. b) Spectral position of the cavity defect, at a temperature of 8 K, as a function of the applied magnetic field (clockwise polarization in red and counter-clockwise polarization in blue).

In Figure 4a we show the light transmission of the microcavity, at 8 K, without magnetic field applied (red curve), and with applied magnetic field of 5 T for the clockwise $\sigma_+$ (blue curve) and the counter-clockwise $\sigma_-$ (dark yellow curve) polarizations. In Figure 4b we show the spectral position of the defect modes as a function of the magnetic field, with a maximum shift of 35 nm at 15 T.

**Conclusion**

In this work we have shown different possibilities of active tuning of the optical response of 1D photonic crystals and microcavities. We review and compare two 1D porous systems: the first one with a liquid crystal infiltrated in the pores of the photonic structure, while the second is a photonic structure employing silver nanoparticles. We also show the light transmission of ITO/titania 1D photonic structures, envisaging the promising electro-optic switching capability. Finally, we describe a magneto-optic in the near infrared made with a TGG based microcavity. Such all-optical switches can be employed in a range of applications, including ICT and electrochromic windows.

*Conflict of Interest*

There are no conflicts of interest to declare.

*Acknowledgement*

G.M.P is supported by the H2020 ETN SYNCHRONICS under a grant agreement 643238. Moreover, This project has received funding from the European Union's Horizon 2020 research and innovation programme (MOPTOPus) under the Marie Skłodowska-Curie grant agreement No. [705444], as well as (SONAR) grant agreement no. [734690]

*References*

1. E. Yablonovitch, "Inhibited Spontaneous Emission in Solid-State Physics and Electronics," Phys. Rev. Lett. **58**(20), 2059–2062 (1987) [doi:10.1103/PhysRevLett.58.2059].
2. S. John, "Strong localization of photons in certain disordered dielectric superlattices," Phys. Rev. Lett. **58**(23), 2486–2489 (1987) [doi:10.1103/PhysRevLett.58.2486].
3. J. D. Joannopoulos, *Photonic crystals: molding the flow of light*, Princeton University Press, Princeton (2008).
4. L. Nucara, F. Greco, and V. Mattoli, "Electrically responsive photonic crystals: a review," J Mater Chem C **3**(33), 8449–8467 (2015) [doi:10.1039/C5TC00773A].
5. A. C. Arsenault et al., "Photonic-crystal full-colour displays," Nat. Photonics **1**(8), 468–472 (2007) [doi:10.1038/nphoton.2007.140].
6. D. P. Puzzo et al., "Electroactive Inverse Opal: A Single Material for All Colors," Angew. Chem. Int. Ed. **48**(5), 943–947 (2009) [doi:10.1002/anie.200804391].
7. A. Baev et al., "Metaphotonics: An emerging field with opportunities and challenges," Phys. Rep. **594**, 1–60 (2015) [doi:10.1016/j.physrep.2015.07.002].
8. D. Bossini et al., "Magnetoplasmonics and Femtosecond Optomagnetism at the Nanoscale," ACS Photonics **3**(8), 1385–1400 (2016) [doi:10.1021/acsphotonics.6b00107].
9. S. Pu, S. Dong, and J. Huang, "Tunable slow light based on magnetic-fluid-infiltrated photonic crystal waveguides," J. Opt. **16**(4), 045102 (2014) [doi:10.1088/2040-8978/16/4/045102].
10. D. Su et al., "A Photonic Crystal Magnetic Field Sensor Using a Shoulder-Coupled Resonant Cavity Infiltrated with Magnetic Fluid," Sensors **16**(12), 2157 (2016) [doi:10.3390/s16122157].
11. I. Kriegel and F. Scotognella, "Magneto-optical switching in microcavities based on a TGG defect sandwiched between periodic and disordered one-dimensional photonic structures," Opt. - Int. J. Light Electron Opt. **142**, 249–255 (2017) [doi:10.1016/j.ijleo.2017.05.091].
12. H. Park et al., "Evaluating liquid crystal properties for use in terahertz devices," Opt. Express **20**(11), 11899 (2012) [doi:10.1364/OE.20.011899].
13. L. Criante and F. Scotognella, "Low-Voltage Tuning in a Nanoparticle/Liquid Crystal Photonic Structure," J. Phys. Chem. C **116**(40), 21572–21576 (2012) [doi:10.1021/jp309061r].
14. L. Criante and F. Scotognella, "Infiltration of E7 Liquid Crystal in a Nanoparticle-Based Multilayer Photonic Crystal: Fabrication and Electro-optical Characterization," Mol. Cryst. Liq. Cryst. **572**(1), 31–39 (2013) [doi:10.1080/15421406.2012.763207].
15. E. Aluicio-Sarduy et al., "Electric field induced structural colour tuning of a silver/titanium dioxide nanoparticle one-dimensional photonic crystal," Beilstein J. Nanotechnol. **7**, 1404–1410 (2016) [doi:10.3762/bjnano.7.131].
16. A. L. Pokrovsky et al., "Theoretical and experimental studies of metal-infiltrated opals," Phys. Rev. B **71**(16) (2005) [doi:10.1103/PhysRevB.71.165114].
17. V. Morandi et al., "Colloidal Photonic Crystals Doped with Gold Nanoparticles: Spectroscopy and Optical Switching Properties," Adv. Funct. Mater. **17**(15), 2779–2786 (2007) [doi:10.1002/adfm.200600764].
18. F. Scotognella et al., "Two-Photon Poly(phenylenevinylene) DFB Laser [†]," Chem. Mater. **23**(3), 805–809 (2011) [doi:10.1021/cm102102w].
19. M. Born and E. Wolf, *Principles of Optics: Electromagnetic Theory of Propagation, Interference and Diffraction of Light*, Cambridge University Press (2000).
20. X. Xiao et al., "Investigation of defect modes with Al2O3 and TiO2 in one-dimensional photonic crystals," Opt. - Int. J. Light Electron Opt. **127**(1), 135–138 (2016) [doi:10.1016/j.ijleo.2015.10.005].

21. M. Bellingeri, I. Kriegel, and F. Scotognella, "One dimensional disordered photonic structures characterized by uniform distributions of clusters," Opt. Mater. **39**, 235–238 (2015) [doi:10.1016/j.optmat.2014.11.033].
22. J. M. Luque-Raigon, J. Halme, and H. Miguez, "Fully stable numerical calculations for finite one-dimensional structures: Mapping the transfer matrix method," J. Quant. Spectrosc. Radiat. Transf. **134**, 9–20 (2014) [doi:10.1016/j.jqsrt.2013.10.007].
23. M. Bellingeri et al., "Optical properties of periodic, quasi-periodic, and disordered one-dimensional photonic structures," Opt. Mater. **72**, 403–421 (2017) [doi:10.1016/j.optmat.2017.06.033].
24. I. Kriegel, F. Scotognella, and L. Manna, "Plasmonic doped semiconductor nanocrystals: Properties, fabrication, applications and perspectives," Phys. Rep. **674**, 1–52 (2017) [doi:10.1016/j.physrep.2017.01.003].
25. G. M. Paternò et al., "Solution processable and optically switchable 1D photonic structures," ArXiv171103192 Cond-Mat Physicsphysics (2017).
26. M. Inoue et al., "Magneto-optical properties of one-dimensional photonic crystals composed of magnetic and dielectric layers," J. Appl. Phys. **83**(11), 6768 (1998) [doi:10.1063/1.367789].
27. V. I. Belotelov and A. K. Zvezdin, "Magneto-optical properties of photonic crystals," J. Opt. Soc. Am. B **22**(1), 286 (2005) [doi:10.1364/JOSAB.22.000286].
28. M. Inoue et al., "Magnetophotonic crystals," J. Phys. Appl. Phys. **39**(8), R151 (2006) [doi:10.1088/0022-3727/39/8/R01].
29. A. M. Grishin and S. I. Khartsev, "All-Garnet Magneto-Optical Photonic Crystals," J. Magn. Soc. Jpn. **32**(2_2), 140–145 (2008) [doi:10.3379/msjmag.32.140].
30. U. Löw et al., "Magnetization, magnetic susceptibility and ESR in Tb3Ga5O12," Eur. Phys. J. B **86**(3) (2013) [doi:10.1140/epjb/e2012-30993-0].
31. R. Yasuhara et al., "Temperature dependence of thermo-optic effects of single-crystal and ceramic TGG," Opt. Express **21**(25), 31443 (2013) [doi:10.1364/OE.21.031443].
32. O. Slezák et al., "Temperature-wavelength dependence of terbium gallium garnet ceramics Verdet constant," Opt. Mater. Express **6**(11), 3683 (2016) [doi:10.1364/OME.6.003683].
33. I. H. Malitson, "Interspecimen Comparison of the Refractive Index of Fused Silica," J. Opt. Soc. Am. **55**(10), 1205–1208 (1965) [doi:10.1364/JOSA.55.001205].
34. S. Fujioka et al., "Kilotesla Magnetic Field due to a Capacitor-Coil Target Driven by High Power Laser," Sci. Rep. **3** (2013) [doi:10.1038/srep01170].
35. F. Mitschke, *Fiber optics: physics and technology*, Springer, Heidelberg ; New York (2009).
36. G. W. C. Kaye and T. H. Laby, *Tables of Physical and Chemical Constants and Some Mathematical Functions (16th edition)* (1995).
37. Y. Nigara, "Measurement of the Optical Constants of Yttrium Oxide," Jpn. J. Appl. Phys. **7**(4), 404–408 (1968) [doi:10.1143/JJAP.7.404].
38. Kruk, Andrzej et al., "Transparent yttrium oxide ceramics as potential optical isolator materials," Opt. Appl. **45**(4), 585–594 (2015) [doi:10.5277/oa150413].